# Generation of entanglement between bright light fields via incoherent spontaneous emission

Xihua Yang[1, 3*], Mingfei Cheng[1, 3], Xiaodong Zeng[1], Zhenghong Li[1], Min Xiao[2*]

[1]*Department of Physics and Institute for Quantum Science and Technology, Shanghai University, Shanghai 200444, China*

[2]*National Laboratory of Solid State Microstructures and School of Physics, Nanjing University, Nanjing 210093, China*

[3]*School of Physics and Technology, Nantong University, Nantong 226019, China*

(Dated: May 1, 2025)

In contrast to the general argument that the spontaneous decay is intrinsically incoherent in nature and detrimental to quantum entanglement, here, we show that nearly perfect entanglement between two bright pump fields can be realized via spontaneous decay-induced destructive quantum interference in a closed double Λ-type four-level atomic system with the energy separation of the excited doublet comparable to their decay rates. The high degree of bipartite entanglement results from the cancellation of spontaneous emission and subsequent elimination of the associated noise due to the destructive quantum interference between the two spontaneous emission pathways from the upper doublet to each of the two lower levels when the two strong pump fields are tuned to the particular frequencies where quantum interference takes place. This scheme is particularly suitable for the generation of entanglement between two bright light fields, and may find potential applications in realistic quantum information processing.

Quantum entanglement provides an essential resource for quantum computation, quantum communication, and quantum information processing. The most conventionally-used tool for generating bipartite entanglement is to employ the parametric down-conversion process in nonlinear crystals, however, such produced entangled fields are usually degenerate and suffer from broad bandwidth and short correlation time, thus limiting their potential applications in realistic quantum memory and quantum networks [1]. To overcome these limitations, the nondegenerate four-wave mixing (FWM) process in a Λ-type atomic system has proven to be an efficient and convenient way for producing nondegenerate and narrow-band entangled fields with long correlation time [2-9]. In order to reduce or eliminate the detrimental effect of noise from spontaneous emission of the excited states on entanglement generation, strong pump fields with one-photon detunings larger than the Doppler broadening are normally employed to interact with atoms [2-4]; an alternative attractive avenue is to use electromagnetically induced transparency (EIT)-based FWM process, which can benefit from the cancellation of resonant absorption and, at the same time, the resonant enhancement of optical nonlinearity [6-9].

It is generally thought that the incoherent process, i.e., spontaneous emission, is dephasing in nature and detrimental to quantum coherence and interference as well as to quantum entanglement. However, the spontaneous emission (SE)-induced quantum effects, such as spontaneous emission cancellation [10-13], spectral line narrowing [14-15], and quantum coherent control of population transfer [16], have been extensively studied. Recently, we have shown that remarkable enhancement of entanglement between the bright pump and probe fields can be achieved via the incoherent collisions in a ladder-type three-level atomic system [17]. Here, we further demonstrate that nearly perfect entanglement between two bright light fields can be realized via the destructive quantum interference between the two spontaneous emission pathways from the upper doublet to each of the two lower levels and subsequent cancellation of spontaneous emission of the excited doublet in a closed double Λ-type four-level system with the energy separation of the excited doublet comparable to their decay rates when the two laser fields are tuned to the particular frequencies where the condition for quantum interference is satisfied. This scheme opens up interesting prospects for generating entanglement between bright light fields.

The considered closed double Λ-type four-level atomic system, as shown in Fig. 1a, interacts with two strong coherent pump fields P1 and P2, where levels 1 and 3 and closely-separated doublet 2 and 4 are coupled by the two pump fields with frequency $\omega_{p1}$ and $\omega_{p2}$, respectively. The detunings of P1 field from the two resonant transitions 1-4 and 1-2 are given by $\Delta_1 = \omega_{p1} - \omega_{41}$ and $\Delta_2 = \omega_{p1} - \omega_{21}$, respectively, where $\omega_{ij}$ (i ≠ j) is the resonant frequency between levels i and j. We assume that the two-photon resonance between the two lower levels 1 and 3 is maintained, so the detunings of P2 field from the two resonant transitions 3-4 and 3-2 are $\omega_{p2} - \omega_{43} = \Delta_1$ and $\omega_{p2} - \omega_{23} = \Delta_2$, respectively. The upper closely-lying levels 2 and 4 are also coupled to the lower levels 1 and 3 by the vacuum modes, and the spontaneous decay rates from the two upper levels to the two lower levels are denoted as $2\gamma_1$, $2\gamma_2$, $2\gamma_3$, and $2\gamma_4$, respectively. The population decay rate between the two lower levels 1 and 3 is denoted as $2\gamma_0$, which may be due to atomic collisions. The quantum operators of the two pump fields and the $k$ th ($q$ th) vacuum mode with frequency $\omega_{k(q)}$ are denoted as $a_1(z,t)$, $a_2(z,t)$, and $a_{k,q}(z,t)$, and the collective atomic operators are denoted as $\sigma_{ij}(z,t)$ (i, j=1, 2, 3, and 4), respectively. The interaction Hamiltonian of the system in the rotating-wave approximation has the form [12, 13,18-20]

$$\hat{V} = -\frac{\hbar N}{L}\int_0^L dz\{\Delta_1\sigma_{44}(z,t)_{44} + \Delta_2\sigma_{22}(z,t) + (\Delta_1 - \Delta_3)\sigma_{33}(z,t) + [g_{14}a_1(z,t)\sigma_{41}(z,t) + g_{12}a_1(z,t)\sigma_{21}(z,t) +$$
$$g_{34}a_2(z,t)\sigma_{43}(z,t) + g_{32}a_2(z,t)\sigma_{23}(z,t) + \sum_k(g_k^{(1)}e^{-i(\omega_k-\omega_{p1})t}a_k(z,t)\sigma_{41}(z,t) + g_k^{(2)}e^{-i(\omega_k-\omega_{p1})t}a_k(z,t)\sigma_{21}(z,t))$$
$$+\sum_q(g_q^{(1)}e^{-i(\omega_q-\omega_{p2})t}a_q(z,t)\sigma_{43}(z,t) + g_q^{(2)}e^{-i(\omega_q-\omega_{p2})t}a_q(z,t)\sigma_{23}(z,t)) + H.c.]\},$$

(1)

where $N$ is the total number of atoms in the interaction volume, $g_{14,12(34,32)} = \mu_{14,12(34,32)} \cdot \varepsilon_{1(2)}/\hbar$ are the atom-field coupling constants with $\mu_{14,12(34,32)}$ being the dipole moments for the 1-4 and 1-2 (3-4 and 3-2) transitions (for simplicity, we assume that the four dipole moments for the transitions 1-4, 1-2, 3-4, and 3-2 are equal to each other) and $\varepsilon_{1(2)} = \sqrt{\hbar \omega_{p1(p2)}/2\epsilon_0 V}$ being the electric field of a single P1 (P2) pump field photon, $\epsilon_0$ is the free space permittivity, $V$ is the interaction volume with length $L$ and beam radius $r$, and $g_{k(q)}^{(1,2)}$ are the coupling constants between the $k$ th ($q$ th) vacuum mode and the atomic transitions from levels 4 and 2 to level 1 (3). Using the Weisskopf-Wigner approximation in the generalized reservoir theory [12, 13, 21], the Heisenberg-Langevin equations for describing the evolutions of the atomic operators and the coupled propagation equations for the two pump field operators can be obtained, which are similar to those in Refs. [12, 13, 16], except that the two fields are treated quantum mechanically instead of classically and the atomic noise terms are taken into account for the present case. In what follows, we will show how to produce nearly perfect entanglement between the two strong pump fields via the spontaneous decay-induced quantum interference in the double Λ-type four-level atomic system.

As is well known, in the traditional Λ-type three-level atomic system (e.g., levels 1, 2 (or 4), and 3 in Fig. 1a) driven by two strong pump fields, when $g_{12(4)}\langle a_1 \rangle$ and $g_{32(4)}\langle a_2 \rangle$ (corresponding to the Rabi frequencies of the two pump fields in semiclassical treatment with $\langle a_1 \rangle$ and $\langle a_2 \rangle$ being the mean values of the two field operators) are much larger than $\sqrt{\gamma_{12(4)}\gamma_{13}}$ (or $\sqrt{\gamma_{32(4)}\gamma_{13}}$), that is, in the coherent population trapping (CPT) regime [22-24], the atoms are pumped into a coherent superposition of the two lower states, i.e., the dark state $|\phi_0\rangle = \cos\theta|1\rangle - \sin\theta|3\rangle$ with $tg\theta = \frac{\langle g_{12(4)}a_1 \rangle}{\langle g_{32(4)}a_2 \rangle}$; owing to the destructive quantum interference between the two transition pathways 1-2 (or 4) and 3-2 (or 4) driven by the two pump fields, the atoms would not absorb the laser radiation, and no population would stay in the excited state 2 (or 4). If one pump field (e.g., P1 field) is far weaker than the other one (e.g., P2 field), the CPT configuration turns into the EIT one, and as discussed in Ref. [8], electromagnetically induced entanglement (EIE) can be realized with a suitable coherent decay rate between the two lower levels. However, nearly no entanglement would exist between the two pump fields in the CPT regime, which is due to the fact that when both $g_{12(4)}\langle a_1 \rangle$ and $g_{32(4)}\langle a_2 \rangle$ are much larger than $\sqrt{\gamma_{12(4)}\gamma_{13}}$ (or $\sqrt{\gamma_{32(4)}\gamma_{13}}$), the atoms will always stay in the dark state $|\phi_0\rangle$ and the medium is almost completely transparent to both pump fields.

In the present double Λ-type four-level system driven by two strong pump fields, as analyzed in Refs. [16, 25], when the two pump fields keep two-photon resonance, but are not tuned to the specific one-photon detuning where the condition for quantum interference is satisfied, i.e., $\mu_{14}^2 \Delta_2 + \mu_{12}^2 \Delta_1 = 0$ and $\mu_{34}^2 \Delta_2 + \mu_{32}^2 \Delta_1 = 0$, there only exists one eigenstate of the Hamiltonian of the atom-field system with zero eigenvalue, which is the dark state $|\phi_0\rangle$. In the strong coupling regime, the atoms would stay in the dark state $|\phi_0\rangle$, just as in the isolated Λ-type three-level system formed by the states 1, 2 (or 4), and 3. In this case, nearly no entanglement would exist between the two strong pump fields. However, when the two laser fields are tuned at the midpoint of the upper doublet, there exist two degenerate eigenstates with the eigenvalues equal to zero; one is the dark state $|\phi_0\rangle$, and the other is

$$|\phi_1\rangle = \sin\theta\sin\varphi|1\rangle + \frac{\cos\varphi}{\sqrt{2}}(|2\rangle - |4\rangle) + \cos\theta\sin\varphi|3\rangle, \qquad (2)$$

where $tg\varphi = \dfrac{\omega_{42}/2}{\sqrt{2(g_{12}^2 a_1^2 + g_{32}^2 a_2^2)}}$, and φ is an additional mixing angle related to the energy separation of the doublet. Due to the nonadiabatic coupling between the two degenerate states $|\phi_0\rangle$ and $|\phi_1\rangle$, the two excited states 2 and 4 would be populated. In this respect, there would exist two emission processes accompanied by the two pump field excitation processes, where one is the SE process, and the other is the stimulated Raman scattering (SRS) process. Similar to the analyzation on EIE in Ref. [8] and collision-induced enhancement of entanglement in Ref. [17], for the case of the SRS processes shown in Fig. 1(b), the two SRS processes together with the two pump field excitation processes, in fact, form a closed-loop light-atom interaction of a FWM process, where the absorption of one photon of P1 (P2) field is always accompanied by the stimulated emission of one photon of P2 (P1) field; subsequently, strong quantum anti-correlation and entanglement between the two pump fields can be realized. However, the SE process is intrinsically incoherent in nature, and the noise from the spontaneous emission of the excited doublet would be detrimental to the entanglement between the two pump fields. The SE and SRS processes would compete with each other, and which process dominates depends on the relative magnitude of the spontaneous decay rates of the excited doublet and the Rabi frequencies of the two pump fields. Therefore, if the spontaneous emission of the excited doublet can be suppressed or cancelled, then the SRS process would become dominated and dramatic enhancement of entanglement generation can be realized. As discussed in Refs. [10-13], the parameters $p_1$ and $p_2$ ( $p_{1(2)} = \dfrac{\vec{\mu}_{21(3)} \cdot \vec{\mu}_{41(3)}}{|\vec{\mu}_{21(3)}| \cdot |\vec{\mu}_{41(3)}|}$ ), which describes the alignment of the two spontaneous emission dipole matrix elements, play a very important role in the quantum interference between the two spontaneous emission pathways from the upper doublet to each of the two lower levels. If the alignment of the two dipole moments is perfectly parallel (or antiparallel), i.e., $p_1 = p_2 = 1$ (or -1), complete destructive (or constructive) quantum interference would take place, and the subsequent spontaneous decay rate can be suppressed (or enhanced); if the alignment of the two dipole moments is orthogonal (i.e., $p_1 = p_2 = 0$ ), there will be no quantum interference. Therefore, one can employ the destructive quantum interference between the two spontaneous emission pathways to suppress or even cancel the spontaneous emission (in view of this, the state $|\phi_1\rangle$ is also a dark state), and subsequently to enhance the entanglement between the two pump fields.

The above prediction is confirmed by solving the Heisenberg-Langevin equations and coupled propagation equations for the interaction of the two pump fields with the atoms. As done in Refs. [8, 17], we use the similar analysis in Ref. [26]

by writing each atomic or field operator as the sum of its mean value and a quantum fluctuation term to treat the atom-field interaction. We consider the case that $g_{12(4)}\langle a_1 \rangle$ and $g_{32(4)}\langle a_2 \rangle$ are much larger than $\sqrt{\gamma_{12(4)}\gamma_{13}}$ (or $\sqrt{\gamma_{32(4)}\gamma_{13}}$), so the depletions of the two pump fields can be safely neglected. The entangled properties of the two pump fields can be tested by the entanglement criterion $V_{12}=(\delta u)^2+(\delta v)^2<4$ proposed in Ref. [27], where $\delta u=\delta x_1+\delta x_2$ and $\delta v=\delta p_1-\delta p_2$ with $\delta x_i=(\delta a_i+\delta a_i^+)$ and $\delta p_i=-i(\delta a_i-\delta a_i^+)$ being the amplitude and phase quadrature fluctuation components of the quantum field operator $a_i$, and the smaller the correlation $V_{12}$ is, the stronger the entanglement gets. It should be noted that the quantum interference of the spontaneous emission pathways can occur only if the transitions from the two closely-separated doublet share the same vacuum modes [10-13], that is, the two levels are close in energy compared to their decay rates. In the following, we set $\omega_{24}=2\gamma_1$ and $\gamma_1=\gamma_2=\gamma_3=\gamma_4=1$ for simplicity, assume the two pump fields to be initially in the coherent states $|a_1\rangle$ and $|a_2\rangle$, and the other parameters are scaled with $\gamma_1$ and m and set according to the realistic experimental conditions [28].

Figure 2 shows the evolution of the correlation $V_{12}$ at zero Fourier frequency and the populations in the four states $\langle\sigma_{11}\rangle$, $\langle\sigma_{22}\rangle$, $\langle\sigma_{33}\rangle$, and $\langle\sigma_{44}\rangle$ as a function of the one-photon detuning $\Delta_1$ while keeping the two-photon resonance satisfied for the case of $p_1=p_2=1$. Obviously, it can be seen from Fig. 2b that when the two pump fields are tuned to the middle point of the two one-photon resonances, the population in the lower state 1 (or 3) appears a narrow dip with a minimal value of about 0.436, whereas the population in the upper state 2 (or 4) appears a narrow peak with a maximal value of about 0.064; in this case, a part of the populations distributed in the two excited states 2 and 4 result from the nonadiabatic coupling between the two degenerate states $|\phi_0\rangle$ and $|\phi_1\rangle$, which would be trapped in the upper doublet due to the complete destructive interference between the two spontaneous emission pathways. Accompanying by the cancellation of the spontaneous emission of the two excited states and subsequent elimination of spontaneous emission noise is the establishment of nearly perfect entanglement between the two strong pump fields through the closed-loop light-atom interaction of the FWM process (see Fig. 2a). Moreover, as displayed in the inset in Fig. 2a, a small deviation of the parameter $p$ ($p = p_1 = p_2$) from unity would lead to the disappearance of the entanglement.

Figure 3 depicts the dependences of the correlation $V_{12}$ at zero Fourier frequency and the absorption coefficients $\alpha_1$ and $\alpha_2$ of the two pump fields on the mean values of the two pump field amplitudes $\langle a \rangle = \langle a_1 \rangle = \langle a_2 \rangle$ for the case of $p_1 = p_2 = 1$ with $\Delta_1 = -\omega_{42}/2$. In order to keep the light-atom interaction in the CPT regime and the intensity absorption rates of the two pump fields nearly stable, as done in Refs. [8, 17], the ratios of $\gamma_{12,13,23}/\alpha_1$ and $n/\alpha_1$ are kept fixed in our calculations. As seen in Fig. 3a, when the intensities of the two pump fields are relatively weak, $V_{12}$ is equal to 4, and no entanglement would exist between the two pump fields. With the increase of $\langle a \rangle$, $V_{12}$ decreases gradually and becomes less than 4, which demonstrates the generation of genuine bipartite entanglement between the two pump fields. When $g_{12(4)} \langle a_1 \rangle$ and $g_{32(4)} \langle a_2 \rangle$ are increased to the order of decay rates of the two excited states, $V_{12}$ becomes nearly equal to zero, which indicates that nearly perfect entanglement between the two pump fields can be realized. However, further increasing $\langle a \rangle$ would weaken and eventually rule out the entanglement. The existence of optimal intensities of two pump fields for generating strongest entanglement is due to the fact that, on one hand, the bipartite entanglement results from the nonlinear interaction (FWM) between the atoms and laser fields, which would be enhanced with increasing the intensities of the two pump fields; on the other hand, as shown in Fig. 3b, the absorption coefficients $\alpha_1$ and $\alpha_2$ of the two pump fields would decrease with the increase of their intensities, exhibiting the analog of CPT in the $\Lambda$-type system, which would trap the atoms in the dark states $|\phi_0\rangle$ and $|\phi_1\rangle$ and weaken the FWM process, and subsequently deteriorate the generation of entanglement.

In Ref. [8], it is shown that the coherence decay rate of the two lower levels plays an essential role in the establishment of entanglement between the pump and probe fields in the $\Lambda$-type three-level EIT configuration, and so it is with the present case. As displayed in Fig. 4a, when there is no dephasing between the two lower levels, $V_{12}$ equals 4, and no entanglement would exist between the two pump fields for the case of $p_1 = p_2 = 1$ with $\Delta_1 = -\omega_{42}/2$. With the increase of the coherence decay rate $\gamma_{13}$, the bipartite entanglement grows strong dramatically, and two nearly perfectly entangled pump fields can be obtained under the conditions of $\gamma_1, \gamma_2 \gg \gamma_{13}$ and $g_{12(4)}\langle a_1\rangle (g_{32(4)}\langle a_1\rangle) \gg \sqrt{\gamma_{12(4)}\gamma_{13}}$ (or $\sqrt{\gamma_{32(4)}\gamma_{13}}$). This is due to the fact that in the strong coupling regime, when there is no population transfer between the two lower levels 1 and 3, i.e., the dephasing rate $\gamma_{13}$ equals zero, the atoms will always stay in the dark states $|\phi_0\rangle$ and $|\phi_1\rangle$, and the medium is completely transparent to both pump fields (see Fig. 4b), so no entanglement can be established between the two pump fields. With the increase of the population transfer between the two lower levels 1 and 3 and the corresponding dephasing rate $\gamma_{13}$, the coherence $\langle \sigma_{ij}\rangle$ $(i \neq j)$ of the atomic system would be weakened; in order to keep the coherence, the two coherent SRS processes would be enhanced accordingly, that is, the closed-loop light-atom interaction of the FWM process as well as the bipartite entanglement between the two pump fields would be strengthened. It should be noted that the dephasing rate $\gamma_{13}$ cannot be increased to the order of the spontaneous decay rates of the two excited states, as in that case, the dissipation process becomes dominant and the approximation of neglecting the depletions of the two pump fields would break down, which would lead to the weakness or even disappearance of entanglement. Similar behavior has been observed for generating pump-probe intensity correlation [29] and squeezed or entangled states of light fields [8, 30, 31] in the Λ-type CPT or EIT configuration.

In order to experimentally observe the above-predicted effects, one can employ the experimental scheme proposed in Ref. [11] with the sodium dimer, where the mixture of the triplet and singlet g-parity Rydberg states by spin-orbit interaction in the sodium dimer can be used to form the two closely-separated levels, and the alignment of the dipole moments between the upper doublet and a lower singlet state (or a triplet state) are parallel (or antiparallel). An alternative convenient way, as discussed in Ref. [21], is to use a third strong coherent pump field to couple the upper level of the Λ-type three-level system to a fourth level to form a double Λ-type four-level system. By suitably tuning the strengths and detunings of the pump fields, the above-mentioned effects from the complete destructive quantum interferences between the spontaneous pathways could be experimentally observed with a suitable coherence decay rate of the two lower levels.

In conclusion, in contrast to the general thought that the spontaneous decay is dephasing in nature and detrimental to entanglement, we demonstrate that nearly perfect entanglement between two bright light fields can be realized via destructive quantum interference between the two spontaneous emission pathways from the upper doublet to each of the lower levels and subsequent cancellation of spontaneous emission of the excited doublet in a closed double Λ-type four-level system when the two laser fields are tuned to the particular frequencies where the condition for quantum interference is satisfied. Our previous [17] and present results clearly indicate that the two incoherent processes, i.e., collision and spontaneous emission, offer a promising alternative to generate entanglement between bright light fields despite of the dephasing nature, which may open up new prospectives for realistic quantum information processing with an atomic or molecular system.

ACKNOWLEDGEMENTS

This work is supported by National Natural Science Foundation of China (Nos. 12174243 and 12474362). Yang's e-mail is yangxh@shu.edu.cn, and Xiao's e-mail is mxiao@uark.edu.

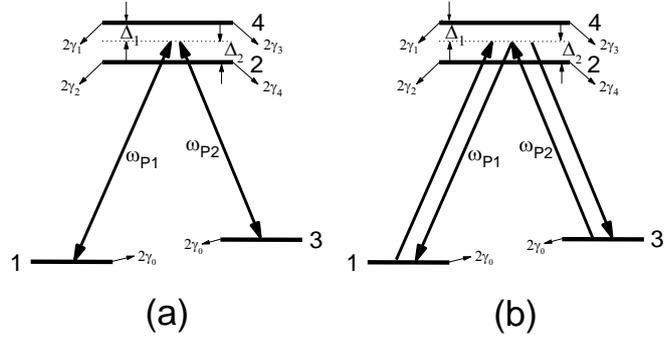

FIG. 1. (**a**) The closed double Λ-type four-level system driven by two strong coherent pump fields with frequency $\omega_{p1}$ and $\omega_{p2}$, respectively. (**b**) The equivalent configuration of (a) for the light-atom interaction forming a closed-loop FWM process.

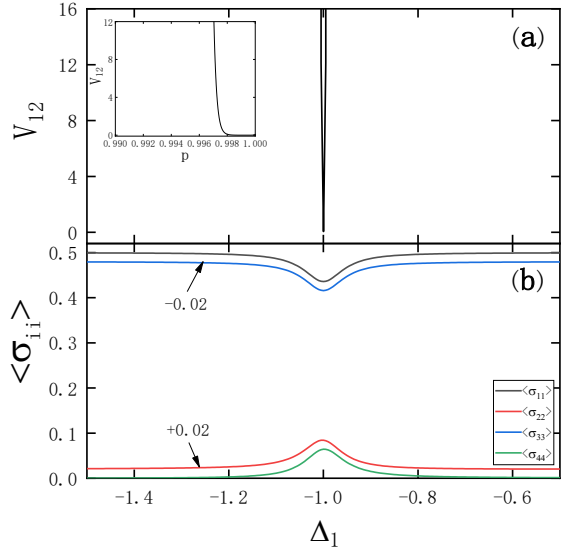

FIG. 2. The dependences of the correlation $V_{12}$ (a) at zero Fourier frequency and the populations (b) of the atom in the four states $\langle\sigma_{11}\rangle$ (dark line), $\langle\sigma_{22}\rangle$ (red line), $\langle\sigma_{33}\rangle$ (blue line), and $\langle\sigma_{44}\rangle$ (green line) on the one-photon detuning $\Delta_1$ with two-photon resonance satisfied with $p_1 = p_2 = 1$, $\omega_{24} = 2\gamma_1$, $\gamma_1 = \gamma_2 = \gamma_3 = \gamma_4 = 1$, $\langle a_1 \rangle = \langle a_2 \rangle = 1$, $r = 2.2\times10^{-4}$, $L$=0.06, $\gamma_0 = 0.001$, and the atomic density $n_0 = 3\times10^{16}$. The inset in Fig. 2a displays the correlation V$_{12}$ at zero Fourier frequency versus the parameter $p$ ($p = p_1$=$p_2$).

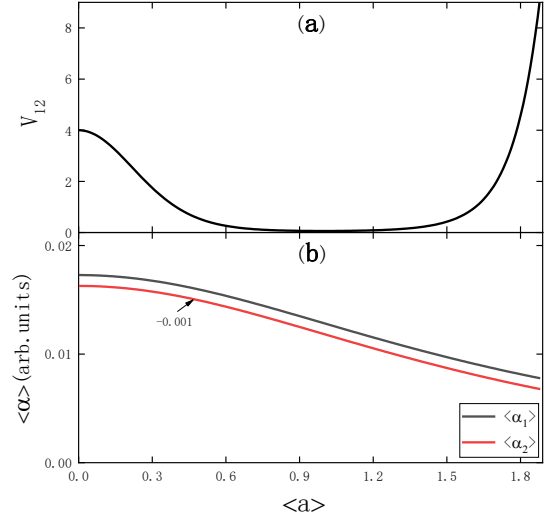

FIG. 3. The dependences of the correlation $V_{12}$ (a) at zero Fourier frequency and the absorption coefficients $\alpha_1$ and $\alpha_2$ (b) of the two pump fields on the mean values of the two pump field amplitudes $\langle a \rangle = \langle a_1 \rangle = \langle a_2 \rangle$ with the atomic density $n = n_0 \langle a \rangle$, $\Delta_1 = -\omega_{42}/2$, $\gamma_0 = 0.001\langle a \rangle$ (a) and $\gamma_0 = 0.001$ (b), and the other parameters are the same as those in Fig. 2.

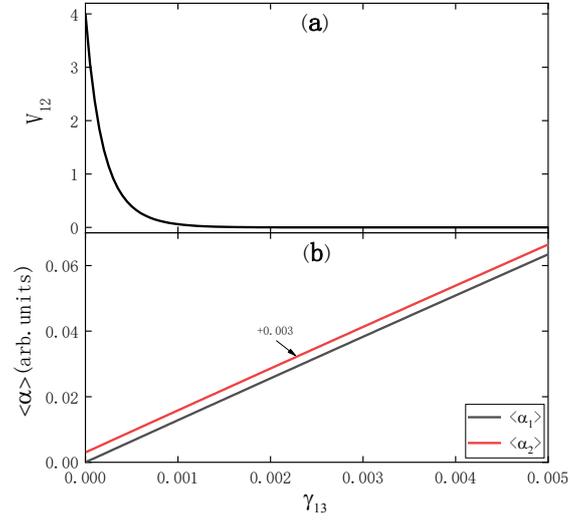

FIG. 4. The dependences of the correlation $V_{12}$ (a) at zero Fourier frequency and the absorption coefficients $\alpha_1$ and $\alpha_2$ (b) of the two pump fields on the coherence decay rate $\gamma_{13}$ with $\Delta_1 = -\omega_{42}/2$, and the other parameters are the same as those in Fig. 2.